\documentclass[12pt,twoside]{article}
\usepackage{cmp2e,amssymb,graphicx}

\title[Thermodynamics of DMAAS and DMAGaS crystals]%
{Thermodynamics and dielectric anomalies of DMAAS and DMAGaS crystals in
the phase transitions region (Landau theory approach)}
\author[I.V.Stasyuk \textit{et al.}]%
{I.V.Stasyuk\refaddr{adr1}, O.V.Velychko\refaddr{adr1},
Z.Czapla\refaddr{adr2}, R.Czukwinski\refaddr{adr2}}
\addresses{
\addr{adr1} Institute for Condensed Matter Physics\\
of the National Academy of Sciences of Ukraine,\\
1 Svientsitskii Str., 79011 Lviv, Ukraine
\addr{adr2} Institute of Experimental Physics, University of Wroclaw,\\
9~M.Borna~Sq., 50--204 Wroclaw, Poland
}
\date{Received December 22, 1999}

\begin{document}

\maketitle

\begin{abstract}
%\tolerance=9999
A simple description of thermodynamics of DMAAS and DMAGaS ferroelectric
crystals by means of Landau expansion is proposed. Conditions of occurrence
of phase transitions are established and their temperatures are obtained. The
influence of external hydrostatic pressure on phase transitions is described.
The temperature behaviour of dielectric susceptibility components and their
anomalies in the vicinity of phase transition points are investigated.
Obtained results are compared with experimental data.
\pacs 77.84.-s, 64.60.Cn
\end{abstract}

%%\documentstyle[12pt,amssymb]{article}
%%\begin{document}

\section{Introduction}

Ferroelectric crystals
(CH$_3$)$_2$NH$_2$Al(SO$_4$)$_2$ $\cdot$ 6H$_2$O (DMAAS) and
\linebreak
(CH$_3$)$_2$NH$_2$Ga(SO$_4$)$_2$ $\cdot$ 6H$_2$O (DMAGaS)
are intensively studied in recent years.
Their interesting feature is possible existence of crystal in ferroelectric
or antiferroelectric state depending on external conditions (e.g.
temperature, hydrostatic pressure). There is a significant difference in
thermodynamical behaviour of crystals despite on isomorphism of their
structure. At ambient pressure DMAGaS crystal has three phases: paraelectric
($T>T_{c}$), ferroelectric ($T_{1}<T<T_{c}$) and antiferroelectric
($T<T_{1}$) with temperatures of phase transitions $T_{c}=136$~K (first
order transition close to the tricritical point) and $T_{1}=117$~K (first
order transition). There is only two phases in DMAAS crystal at
ambient pressure:  paraelectric ($T>T_{c}$) and ferroelectric ($T<T_{c}$)
with $T_{c}=155$~K.

A set of structural \cite{r1,r2,r3}, dilatometric, dielectric, pyroelectric
and ultrasonic \cite{r4,r5,r6,r7,r8,r9,r10,r11,r12}  measurements is made
for considered systems, what allows
to establish their main dielectric, mechanical and dynamical
characteristics (see below). At the same time these investigations are
incomplete and of preliminary stage in many directions.

The nature of phase transitions in DMAAS and DMAGaS crystals was unclear up
to recent time. During the last years conviction on important role of
dimethyl ammonium (DMA) groups in phase transitions due to their
orientational ordering-disordering is established (see, for example,
\cite{r3,r6,r13,r14,a2,b1}). In \cite{r15} the microscopic approach based on
the
order-disorder model with account of different orientational states of DMA
groups was proposed. In the framework of the model the phase transition to
ferroelectric state has been described and conditions of realization of this
transition as of the first or of the second order have been established.
Order parameters of the system have been constructed.  They are connected
with differences of occupancies of four possible positions of nitrogen ions
corresponding to different orientations of groups. As a result of symmetry
analysis it has been established that components of the order parameters
belonging to irreducible representation $B_{u}$ of point symmetry group
$2/m$ of the high-temperature (paraelectric) phase describe ferroelectric
ordering of DMA group along the ferroelectric axis OX (in crystallographic
plane (ac)) and their antiferroelectric ordering along the OY axis
(crystallographic axis b). The inverse ordering (antiferroelectric along OX
and ferroelectric one along OY) corresponds to order parameter components
belonging to irreducible representation $A_{u}$.  Appearance of nonzero
order parameters of $B_{u}$ type turns the system into ferroelectric state
(point group $m$) while nonzero order parameters of $A_{u}$ type cause
antiferroelectric state (point group 2).

Notwithstanding further perspectives of microscopic approach by means of the
four-state order-disorder model, the more simple but more general
thermodynamical description based on Landau expansion is of interest.  One
can construct corresponding Landau free energy and in standard way
investigate possible phase transitions and obtain criteria of their
realization with the use of data of the mentioned above symmetry analysis.
This is a main goal of the present work. Results obtained in the framework
of Landau expansion will be used for interpretation of the induced by the
external pressure changes in the picture of phase transitions and for
description of dielectric anomalies in the phase transition points of the
investigated crystals.

\section{Thermodynamics of phase transitions (Landau theory 
\protect\linebreak
approach)}

Let us make thermodynamical description of phase transition in DMAAS and
DMAGaS crystals with the use of Landau expansion. We consider a simplified
version when only one linear combination of the initial order parameters
type is included for each of $B_{u}$ and $A_{u}$ irreducible
representations. The combinations included are true order parameters:
coefficients at their squared values tend to zero in the points of
corresponding second order transitions.

Order parameters, which transform according to irreducible representations
$B_{u}$ and $A_{u}$ of point symmetry group $2/m$ of high-symmetry phase,
are denoted as $\eta_{b}$ and $\eta_{a}$ correspondingly. The first
parameter $\eta_{b}$ describes polarization of ferroelectric type along the
OX axis with simultaneous antiferroelectric type ordering along the OY axis;
the second one corresponds to inverse orientation where antipolarization
along OX is accompanied by polarization along OY.

We restrict ourself to the case of second order phase transition from the
nonpolar high-temperature phase to ordered one. In this case Landau
expansion of free energy can be limited by terms of the fourth order:
\begin{equation}
F=F_{0}+\frac{1}{2} a\eta_{a}^{2} + \frac{1}{2} b\eta_{b}^{2} + \frac{1}{4}
c \eta_{a}^{4} + \frac{1}{4} d\eta_{b}^{4} + \frac{1}{2}
f\eta_{a}^{2}\eta_{b}^{2} - E_{x}\eta_{b} - E_{y}\eta_{a} \: .
\label{eq1}
\end{equation}
A linear dependence of coefficients $a$ and $b$ on temperature is assumed
\begin{equation}
a=a'(T-T_{c}'), \qquad b=b'(T-T_{c}),
\label{eq2}
\end{equation}
where condition $T_{c}>T_{c}'$ is satisfied for normal state of the
crystal what corresponds to the transition from the paraelectric phase (phase P)
to the ferroelectric phase (phase F) as to the first one at lowering of
the temperature.

Conditions of thermodynamical equilibrium correspond to the minimum of free
energy and look like
\begin{eqnarray}
\frac{\partial F}{\partial\eta_{a}} &=& \eta_{a} (a+c\eta_{a}^{2} +
f\eta_{b}^{2}) - E_{y} = 0, \nonumber\\
\frac{\partial F}{\partial\eta_{b}} &=& \eta_{b} (b+c\eta_{b}^{2} +
f\eta_{a}^{2}) - E_{x} = 0.
\label{eq3}
\end{eqnarray}
At zero external fields there are following solutions
\begin{equation}
\eta_{a}=\eta_{b}=0
\label{eq4}
\end{equation}
-- paraphase (P-phase);
\begin{eqnarray}
&&\eta_{a} = 0, \quad \eta_{b}\neq 0, \nonumber\\
&&\eta_{b0} = \sqrt{-b/d} = \sqrt{(b'/d)(T_{c}-T)}
\label{eq5}
\end{eqnarray}
-- ferroelectric phase (F-phase);
\begin{eqnarray}
&&\eta_{a}\neq 0; \quad \eta_{b}=0 \nonumber\\
&&\eta_{a0} =\sqrt{-a/c} = \sqrt{a'/c (T_{c}'-T)}
\label{eq6}
\end{eqnarray}
-- antiferroelectric phase (AF-phase).
\footnote{We follow here to the terminology widely used in literature on the
subject.}

\begin{figure}
\centerline{\includegraphics[width=0.8\textwidth]{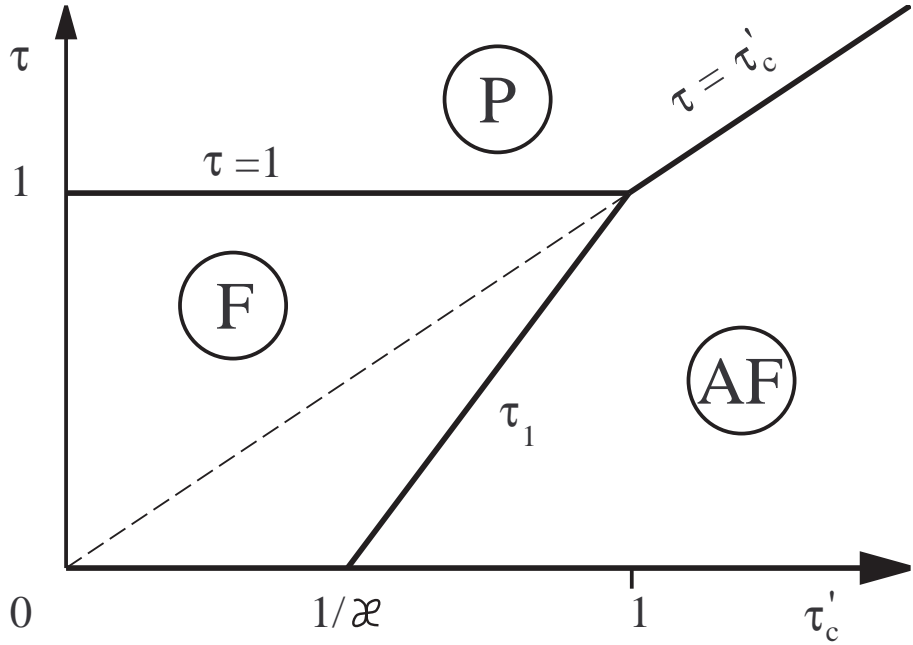}}
\caption{Dependence of phase number of the system and phase transition
temperatures on values of system parameters}
\label{lndfig1}
\end{figure}

Corresponding expressions for free energy in these phases are as follows
\begin{eqnarray}
F_{(P)} &=& F_{0}\:, \nonumber\\
F_{(F)} &=& F_{0} - \frac{1}{4d} b'^{2} (T-T_{c})^{2}, \\
F_{(AF)} &=& F_{0} - \frac{1}{4c} a'^{2} (T-T_{c}')^{2}. \nonumber
\label{eq7}
\end{eqnarray}
The phase transition P$\rightarrow$F which is of the second order in the
used approximation takes place at temperature $T_{c}$. The phase transition
F$\rightarrow$AF which can take place at lower temperatures occurs at
\begin{equation}
F_{(F)} = F_{(AF)}.
\label{eq8}
\end{equation}
The condition above determines the temperature
of this first order phase transition:
\begin{equation}
T_{1}=\frac{\varkappa T_{c}'-T_{c}}{\varkappa-1},
\label{eq9}
\end{equation}
where
\begin{equation}
\varkappa=\frac{a'\sqrt{d}}{b'\sqrt{c}}, \qquad \varkappa > 1.
\label{eq10}
\end{equation}
Nonequalities
\begin{equation}
0<T_{1}<T_{c}
\label{eq11}
\end{equation}
define the region of temperature $T_{c}'$ values where the F-phase exists as
an intermediate one:
\begin{equation}
\frac{1}{\varkappa} < \frac{T_{c}'}{T_{c}} <1.
\label{eq12}
\end{equation}
These conditions are illustrated by the phase diagram in Fig.~\ref{lndfig1}.
In the case $T_{c}' > T_{c}$ a direct phase transition P$\rightarrow$AF from
the paraelectric phase to antiferroelectric one can take place.

Observed by experiment changes of temperatures of P$\rightarrow$F and
F$\rightarrow$AF phase transitions and consecutive disappearance of the
F-phase as the result of increasing of external hydrostatic pressure can be
easily explained with the use of the obtained diagram. Under assumption that the
influence of pressure leads mainly to shifts of temperatures $T_{c}$ and
$T_{c}'$
\begin{eqnarray}
T_{c} &=& T_{c0} +x p, \nonumber\\
T_{c}' &=& T_{c0}' + x' p,
\label{eq13}
\end{eqnarray}
and the changes of other Landau expansion parameters are negligible, the
following relation is obtained
\begin{equation}
T_{1} = T_{1}^{0} + \frac{\varkappa x' - \varkappa}{\varkappa-1}p,
\label{eq14}
\end{equation}
where
\begin{equation}
T_{1}^{0} = \frac{\varkappa T_{c0}' - T_{c0}}{\varkappa-1}.
\label{eq15}
\end{equation}
According to the data published in \cite{r16}, $dT_{c}/dp \equiv x=-0.277$~K/MPa;
$\partial T_{1}/\partial p= 1.95$~K/MPa and if one applies a linear
approximation to the dependence of $T_{c}'$ on $p$ then
$dT_{c}'/dp \equiv x'= 0.86$~K/MPa.

\begin{figure}
\centerline{\includegraphics[width=0.8\textwidth]{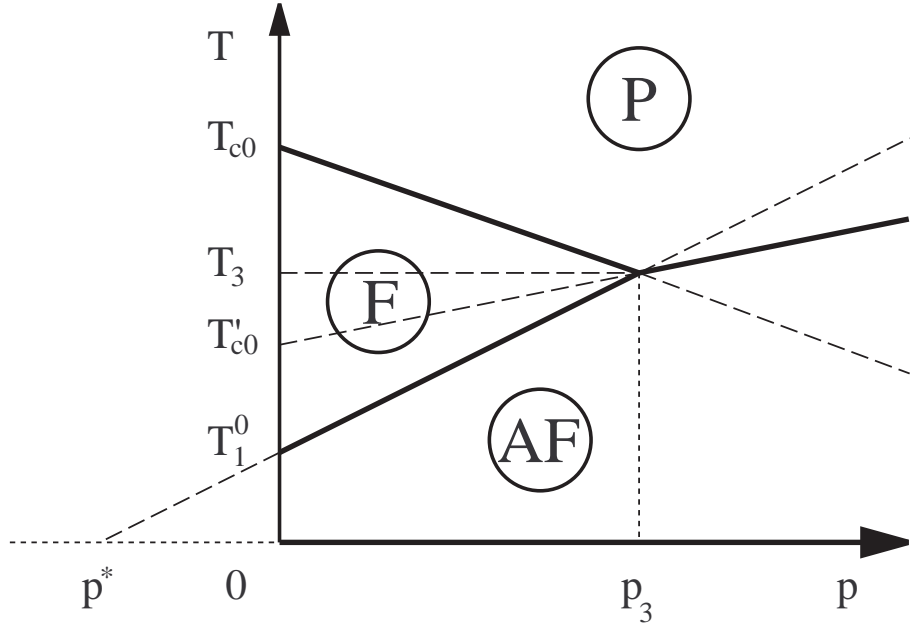}}
\caption{Dependence of the phase transition
temperatures on the external hydrostatic pressure}
\label{lndfig2}
\end{figure}

The obtained relations are illustrated by the diagram shown in Fig.~2. This
diagram qualitatively matches the experimental (T,p) diagram for DMAGaS
crystal (at $T_{c0}=136$~K, $T_{10}=116$~K) \cite{r16}. Obtained coordinates of
triple point
\begin{equation}
T_{3}=\frac{x'T_{c0}-xT_{c0}'}{x'-x}, \qquad
P_{s}=\frac{T_{c0}-T_{c0}'}{x'-x},
\label{eq17}
\end{equation}
where lines of phase transitions P$\rightarrow$F, F$\rightarrow$AF and
P$\rightarrow$AF come together  are in good agreement with experimental ones
($T_{3}^{exp}=-140.3$~$^{\circ}$C; $P_{3}^{exp}=8.75$~MPa).
At $p\gg{}p_{3}$ there take place a deviation of the
theoretical prediction of temperature of the P$\rightarrow$AF phase
transition from experimental data. Unlike to relationship
(\ref{eq13}) experimental dependence is nonlinear at large pressures.

The pressure value
\begin{equation}
p^{*} = \frac{\varkappa T_{c0}' - T_{c0}}{x-\varkappa x'}
\label{eq18}
\end{equation}
(see Fig.~\ref{lndfig2}) is an important characteristic of the model.
At $p^{*}<0$, what is realized at
$T_{c0}'/T_{c0}>1/\varkappa$, AF-phase exists in the region of low
temperatures at ambient pressure (this situation takes place for DMAGaS). At
$p^{*}>0$ (i.e.\ $T_{c0}'/T_{c0}<1/\varkappa$) and ambient pressure only P-
and F-phases occur; this case can correspond to DMAAS crystal.

\section{Dielectric susceptibility}

The approach used in the previous section allows to derive expressions for
components of dielectric susceptibility tensor in the vicinity of phase
transition points and to describe their temperature dependencies in general.
In the used approximation the components $P_{x}$ and $P_{y}$ of polarization
vector are defined by parameters $\eta_{b}$ and $\eta_{a}$ correspondingly.
Hence
\begin{equation}
\chi_{xx} = \frac{\partial\eta_{b}}{\partial E_{x}}, \qquad
\chi_{yy} = \frac{\partial\eta_{a}}{\partial E_{y}}
\label{eq19}
\end{equation}
and proceeding from equations (\ref{eq3}) one can obtain
\begin{eqnarray}
\chi_{xx} &=& \frac{1}{D} (a+3c\eta_{a}^{2} + f\eta_{b}^{2}), \nonumber\\
\chi_{yy} &=& \frac{1}{D} (b+3d\eta_{b}^{2} + f\eta_{a}^{2}),
\label{eq20}
\end{eqnarray}
where
\begin{equation}
D=(a+3c\eta_{a}^{2} + f\eta_{b}^{2}) (b+ 3d\eta_{b}^{2} + f\eta_{a}^{2}) -4
f^{2} \eta_{a}^{2} \eta_{b}^{2}\:.
\label{eq21}
\end{equation}
The following particular cases follow from expression (\ref{eq20}):
\begin{enumerate}
\item Paraphase (P):
\begin{equation}
\chi_{xx} = \frac{1}{b} = \frac{1}{b'(T-T_{c})}, \quad
\chi_{yy}=\frac{1}{a}=\frac{1}{a'(T-T_{c}')}.
\label{eq22}
\end{equation}
\item Ferroelectric phase (F):
\begin{equation}
\chi_{xx} = -\frac{1}{2b} = \frac{1}{2b'(T_{c}-T)}, \quad
\chi_{yy}=\frac{1}{(\xi-1) a' (T^{*}-T)},
\label{eq23}
\end{equation}
here the notations are used:
\begin{equation}
T^{*} = T_{c} + \frac{T_{c} - T_{c}'}{\xi - 1}, \quad \xi = \frac{fb'}{da'}
\quad (\xi >1).
\label{eq24}
\end{equation}
In this case susceptibility $\chi_{yy}$ can be also expressed in the form
\begin{equation}
\chi_{yy} = [a+f\eta_{b0}^{2}]^{-1},
\label{eq25}
\end{equation}
where $\eta_{b0}$ is a spontaneous value of order parameter (polarization
$P_{s}$) in the ferroelectric phase.
\item Antiferroelectric phase (AF):
\begin{equation}
\chi_{xx}=\frac{1} {(\varkappa^{2}\xi -1)b' (T^{**}-T)}, \quad
\chi_{yy} =-\frac{1}{2a} = \frac{1}{2a'(T_{c}' - T)},
\label{eq26}
\end{equation}
where the temperature
\begin{equation}
T^{**} = T_{c}+ \frac{T_{c}-T_{c}'}{1-1/\varkappa^{2}\xi},
\label{eq27}
\end{equation}
is introduced such that $T^{**}>T^{*}>T_{c}$. A similar to the previous
one expression
\begin{equation}
\chi_{xx} = [b+f\eta_{a0}^{2}]^{-1},
\label{eq28}
\end{equation}
relating the temperature dependence of longitudinal
susceptibility in AF phase with the equilibrium value of the order parameter
(polarization in one of sublattices) takes place.
\end{enumerate}

\begin{figure}
\centerline{\includegraphics[width=0.8\textwidth]{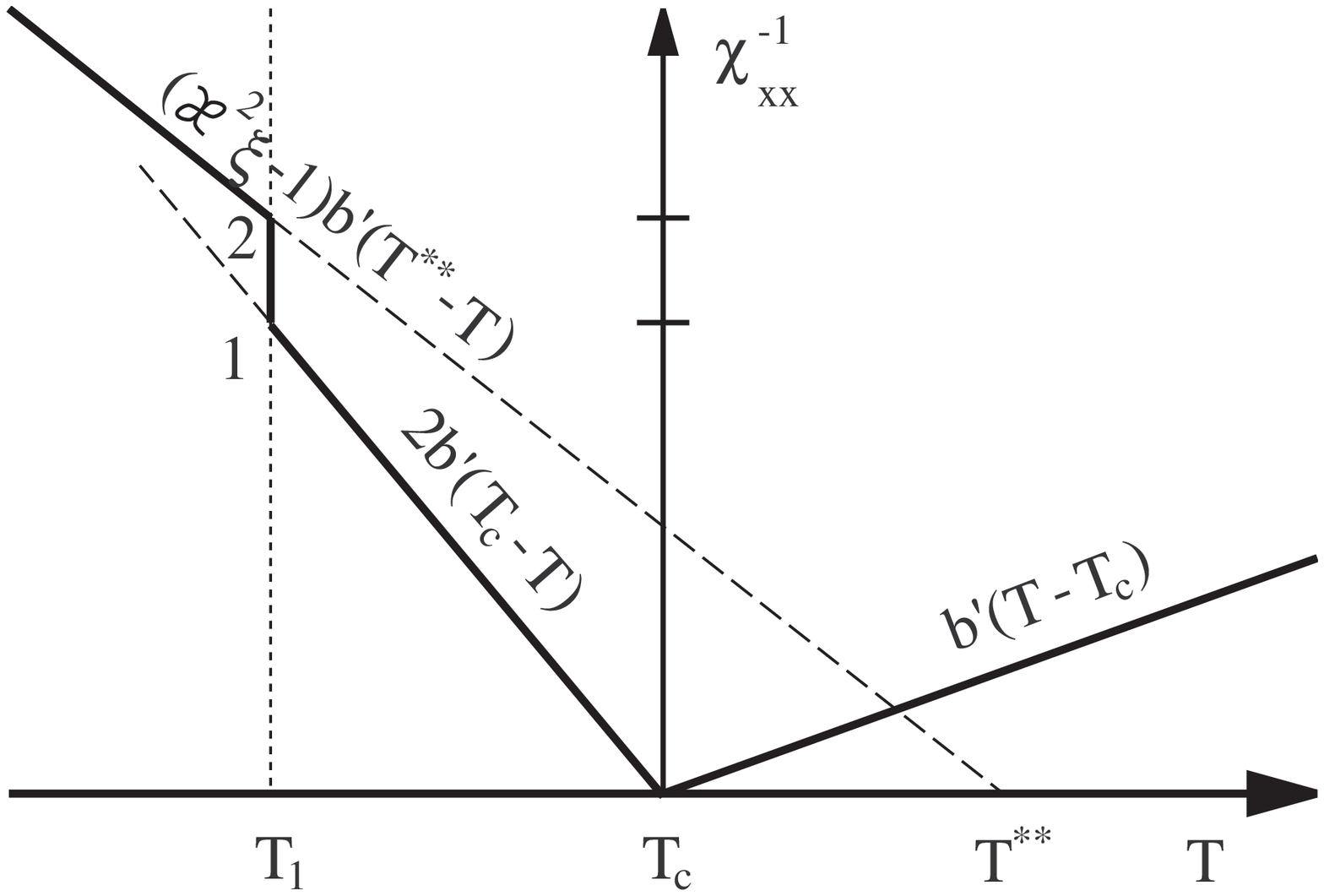}}
\caption{Temperature dependences of $\chi_{xx}^{-1}$ component
of inverse susceptibility.}
\label{lndfig3}
\vspace{7ex}

\centerline{\includegraphics[width=0.8\textwidth]{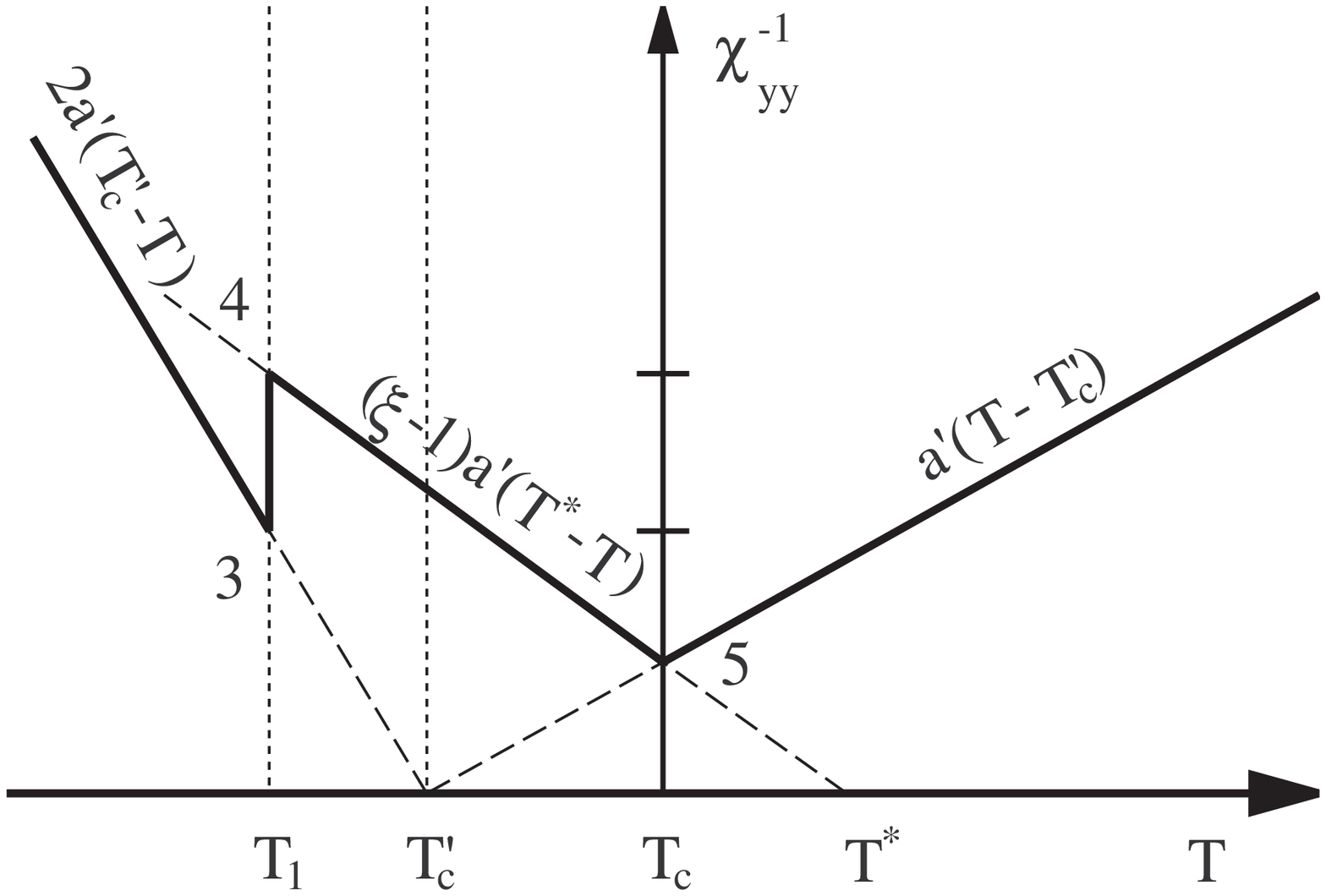}}
\caption{Temperature dependences of $\chi_{yy}^{-1}$ component
of inverse susceptibility.}
\label{lndfig4}
\end{figure}

The temperature behaviour of dielectric susceptibility components and their
anomalies in the phase transition points are illustrated in
Fig.~\ref{lndfig3} and \ref{lndfig4} as temperature dependencies of inverse
susceptibilities $\chi_{\alpha\alpha}^{-1}$.

The inverse susceptibility $\chi_{xx}^{-1}$ is equal to zero at the
temperature $T_{c}$. Its linear dependence on temperature in the vicinity of
this point has an inclination $b'$ at $T>T_{c}$ and $2b'$ in the ferroelectric
phase (Fig.~\ref{lndfig3}). This typical behaviour for second order phase
transition changes if the phase transition P$\rightarrow$F is of the first
order. Such a situation takes place in the DMAGaS crystal where the first
order phase transition close to the tricritical point is observed. Then
the susceptibility $\chi_{xx}^{-1}$ remains nonzero at $T_{c}$ and has a
small jump (according to data \cite{r11}, $T_{c}-T_{0}\simeq1.2$~K,
where $T_{0}$ is the temperature at which $\chi_{xx}^{-1}\rightarrow 0$;
$\chi_{xx}^{-1}(T=T_{c})\simeq6\cdot10^{-4}$).
Mentioned changes are relevant only to a small vicinity of
$T_{c}$; in a large temperature scale dependence $\chi_{xx}^{-1}(T)$ in
para- and ferroelectric phase is almost the same as for the second order
transition. The phase transition F$\rightarrow$AF is a well pronounced first
order phase transition accompanied by jump of the $\chi_{xx}^{-1}$
function. The continuation of the straight line describing the temperature
dependence of $\chi_{xx}^{-1}$ in the AF phase passes the point
$T^{**}$ (see Fig.~3). $\chi_{xx}^{-1}$ has the following values at the
ends of its jump
\begin{eqnarray}
\left.\chi_{xx}^{-1}\right|_{1} &=& 2b' \frac{\varkappa}{\varkappa-1}
(T_{c} - T_{c}'), \nonumber\\
\left.\chi_{xx}^{-1}\right|_{2} &=& \left[ \frac{\varkappa}{\varkappa-1}
(\varkappa^{2}\xi-1) + \varkappa^{2}\xi\right] b' (T_{c}-T_{c}').
\label{eq29}
\end{eqnarray}
Value of susceptibility jump $\Delta \chi_{xx}^{-1} = \left.\chi_{xx}^{-1}
\right|_{2} - \left. \chi_{xx}^{-1}\right|_{1}$ can be positive or negative
depending on values of theory parameters.

Temperature behaviour of the inverse susceptibility $\chi_{yy}^{-1}$ is
essentially different. In the point of the second order phase transition
P$\rightarrow$F it remains nonzero with value
\begin{equation}
\chi_{yy}^{-1} (T_{c}) = a' (T_{c} - T_{c}')
\label{eq30}
\end{equation}
Its continuation to lower temperatures goes to zero at
$T\rightarrow T_{c}'$. The continuation of the line of the inverse
susceptibility in the antiferroelectric phase $\chi_{yy}^{-1}(T) =
2a'(T_{c}'-T)$ also goes across this point. In the ferroelectric
phase region
the function $\chi_{yy}^{-1}(T)$ is linear with the continuation passing
the point $T^{**}$. At the F$\rightarrow$AF phase transition this
function has a jump between points
\begin{eqnarray}
\left. \chi_{yy}^{-1}\right|_{3} &=& \frac{2a'}{\varkappa-1} (T_{c}-T_{c}'),
\nonumber\\
\left. \chi_{yy}^{-1}\right|_{4} &=& \frac{\varkappa\xi -1}{\varkappa-1} a'
(T_{c}-T_{c}').
\label{eq31}
\end{eqnarray}
Similarly to the case of the function $\chi_{xx}^{-1}$ the jump can have
positive or negative value.

\section{Discussion}

Proceeding from obtained in the previous section formulae one can try to
interpret available data on the temperature dependence of dielectric
susceptibility components of DMAAS and DMAGaS crystals. The majority of
performed measurements is devoted to the longitudinal dielectric permittivity
$\varepsilon_{x}$ (or its real part $\varepsilon_{x}'$ for low frequency
alternating current measurements) mainly in the region of the
high-temperature phase transition for DMAGaS and the corresponding phase
transition in DMAAS.  Such data are reported in works \cite{r5,r8,r9,r11}
(DMAGaS) and \cite{r4,r7,r8} (DMAAS);
only in paper \cite{r7} the temperature behaviour of all permittivity
components ($\varepsilon_{a}'$, $\varepsilon_{b}'$, $\varepsilon_{c}'$) for
DMAAS crystal in the wide range of temperatures (from $\simeq$90~K to
$\simeq$280~K) was measured. In some papers dependence of spontaneous
polarization on temperature in the ferroelectric phase was investigated and
coercivity fields were measured \cite{r8,r9} (the value of $P_{s}$
in the state close to
saturation is about 1.4--1.9~C/m$^{2}$ for DMAAS and
0.9--2.0~C/m$^{2}$ for DMAGaS).
Particular investigation of the $T_{c}$ point vicinity in DMAGaS devoted to
influence of the external electric field on the first order phase transition point and
the difference $T_{c}-T_{0}$ is made in \cite{r11}. On the basis of
available experimental data Curie-Weiss constant (from the paraphase side)
is estimated as 2700--3060~K for DMAGaS crystal and 2700--3000~K for DMAAS
crystal. The phase transition to the ferroelectric phase in DMAGaS crystal
is of the first order and close to the tricritical point;
this fact however does not affect the behaviour of
$\chi_{xx}$ and $\chi_{yy}$ far from the $T_{c}$ point.

The mentioned experimental data are incomplete, hence only partial
comparison with results of thermodynamical  description is possible. For
example one can obtain values of the temperature $T_{c}'$, parameters $b'$
and $\varkappa$ for the DMAGaS crystal  $T_{c}'=125$~K, $\varkappa=2.22$,
$b'=0.33\cdot10^{-3}$~K$^{-1}$ with use of above mentioned data on the
influence of external hydrostatic pressure on phase transitions in DMAGaS
crystal \cite{r16} and results of measurements of dielectric
characteristics.

More comprehensive and selfconsistent evaluation of temperatures $T_{c}'$,
$T^{*}$ and $T^{**}$ as well as Landau expansion parameters (or parameters
$a'$, $b'$, $\varkappa$, $\xi$ and $f$) by means of presented in this section
relationships become possible after goal-oriented investigations of
temperature dependencies of $\chi_{xx}^{-1}$ and $\chi_{yy}^{-1}$ in a wide
temperature interval including regions of existence of all phases for DMAGaS
and DMAAS. Proceeding from obtained results will be possible to ascertain
suitability of the simple thermodynamical description where Landau expansion
is limited to only one order parameter for each of $B_{u}$ and $A_{u}$
representations. Such a description is obviously much simplified
comparatively to results of the microscopic approach based on the four-state
model of order-disorder type \cite{r15}. Investigation of DMA group ordering in the
configurational space of four orientational states needs two-component order
parameters $\eta_b^{\alpha}$ ($B_{u}$) and $\eta_a^{\alpha}$ ($A_{u}$),
$\alpha=1,2$. This fact
could complicate temperature dependencies of dielectric characteristics of
the model even for thermodynamical description in the framework of Landau
expansion.

Furthermore, considered here expression for Landau expansion of free energy
(\ref{eq1}) includes terms up to the fourth order. A consistent description
of the first order phase transition P$\rightarrow$F and related dielectric
anomalies demands the inclusion of the sixth order terms into expansion.
Such a generalization is necessary for comprehensive description of
experimental data and can be performed relatively easy.

\label{last@page}

\end{document}